\documentclass[aps,pre,reprint,superscriptaddress,amsfonts,amssymb,amsmath,onecolumn,notitlepage,12pt]{revtex4-1}

\usepackage{bm}
\usepackage[pdftex]{graphicx}
\usepackage[pdftex]{hyperref}

\newcommand{\Mem}{\text{M}}
\newcommand{\Sys}{\text{S}}
\newcommand{\Bath}{\text{B}}
\newcommand{\init}{\text{init}}

\newcommand{\can}{\text{can}}
\newcommand{\meas}{\text{meas}}
\newcommand{\eras}{\text{eras}}
\newcommand{\all}{\text{all}}
\newcommand{\tot}{\text{tot}}

\newcommand{\idop}{\Hat{I}}
\newcommand{\Hami}{\Hat{H}}
\newcommand{\hU}{\Hat{U}}
\newcommand{\hP}[2]{\Hat{P}^{#1}_{#2}}
\newcommand{\hPi}[2]{\Hat{\Pi}^{#1}_{#2}}
\newcommand{\hQ}[2]{\Hat{Q}^{#1}_{#2}}
\newcommand{\hrho}[2]{\Hat{\rho}^{#1}_{#2}}

\newcommand{\EM}[1]{E^\Mem_{#1}}
\newcommand{\EB}[1]{E^\Bath_{#1}}
\newcommand{\HM}{\Hami^\Mem}
\newcommand{\HS}{\Hami^\Sys}
\newcommand{\HB}{\Hami^\Bath}
\newcommand{\HSM}{\Hami^{\Sys\Mem}}
\newcommand{\HMB}{\Hami^{\Mem\Bath}}
\newcommand{\hMB}{\Hat{h}^{\Mem\Bath}}
\newcommand{\rhoS}[1]{\hrho{\Sys}{#1}}
\newcommand{\rhoM}[1]{\hrho{\Mem}{#1}}
\newcommand{\rhoB}[1]{\hrho{\Bath}{#1}}
\newcommand{\rhoSMB}[1]{\hrho{\Sys\Mem\Bath}{#1}}
\newcommand{\rhoMB}[1]{\hrho{\Mem\Bath}{#1}}
\newcommand{\PM}[1]{\hP{\Mem}{#1}}
\newcommand{\PB}[1]{\hP{\Bath}{#1}}
\newcommand{\QS}[1]{\hQ{\Sys}{#1}}

\newcommand{\Wm}{W_\meas}
\newcommand{\We}{W_\eras}
\newcommand{\Fm}{F_\meas}
\newcommand{\Fe}{F_\eras}
\newcommand{\Hm}{H_\meas}
\newcommand{\He}{H_\eras}

\newcommand{\lTr}[1]{\text{Tr} \! \left[ #1 \right]}
\newcommand{\nTr}[1]{\text{Tr} \: #1}
\newcommand{\Tr}[1]{\text{Tr} \! \left( #1 \right)}
\newcommand{\sTr}[2]{\text{Tr}_{#1} \! \left( #2 \right)}
\newcommand{\avr}[1]{\left\langle #1 \right\rangle}
\newcommand{\mavr}[1]{\avr{#1}_\text{m}}
\newcommand{\eavr}[1]{\avr{#1}_\text{e}}
\newcommand{\ket}[1]{\left| #1 \right\rangle}
\newcommand{\bra}[1]{\left\langle #1 \right|}
\newcommand{\ketbra}[1]{\ket{#1}\!\bra{#1}}
\newcommand{\set}[1]{\big\{ #1 \big\}}
\newcommand{\dg}{\dagger}
\newcommand{\Ox}{\otimes}

\newcommand{\Step}[1]{\paragraph*{Step #1}}
\newcommand{\affA}{
Department of Physics, The University of Tokyo, Komaba, Meguro, Tokyo 153-8505 
}
\newcommand{\affB}{Centre for Quantum Technology, National University of Singapore,
Singapore 117543}

\begin{document}
\title{Quantum Jarzynski equalities for the energy costs of the information processes}
\author{Yohei Morikuni}
\affiliation{\affA}
\author{Hiroyasu Tajima}
\affiliation{\affA}
\affiliation{\affB}
\date{\today}

\begin{abstract}
We present quantum versions of the Jarzynski equality for the energy costs of information processes, namely the measurement and the information erasure.
We also obtain inequalities for the energy costs of the information processes, using the Jensen inequality.
The inequalities include Sagawa and Ueda's inequalities~\cite{Sagawa2009} as a special case.
\end{abstract}

\maketitle

\section{Introduction} 
In 1867, Maxwell summoned his famous demon and pointed out that the second law of thermodynamics is seemingly violated in the thermodynamic processes with measurements and feedback controls.
Since then, thermodynamics with information processes have been the center of attention and numerous studies have been done~\cite{Maxwell1871,Thomson1874,Leff2003,Landauer1961,Shizume1995,Sagawa2008,Maroney2009,Sagawa2009,Sagawa2010,Ponmurugan2010,Horowitz2010,Horowitz2011,Sagawa2012,Toyabe2010,Morikuni2011,Rana2012,Ito2013,Tasaki2013,Funo2013,Tajima2013,Tajima2013a}.

The researches for such thermodynamic processes mainly consist of two types.
  The first type of researches considers thermodynamic systems and tries to solve the seeming violation of the second law~\cite{Sagawa2008,Tajima2013}.
  In the researches of  this type, we measure the excess of the work extracted from a thermodynamic system over the conventional second law.
   The excess is equal to the amount of correlation extracted by the measurement; the correlation is measured by the classical mutual information in classical systems and by entanglement of formation in quantum systems~\cite{Sagawa2008,Tajima2013}.
  The second type of researches considers the energy costs of the information processes and seeks the restoration of the second law~\cite{Landauer1961,Shizume1995,Maroney2009,Sagawa2009,Tajima2013a}.
  In the researches of this type, we measure the excess of the energy costs of the information processes such as measurement and information erasure and compare the energy costs with the excess which is derived in the first type.
  
  For classical systems, the above two types have been recently integrated into the form of the fluctuation theorem~\cite{Sagawa2010,Ponmurugan2010,Horowitz2010,Horowitz2011,Sagawa2012,Ito2013,Tasaki2013}; 
  we can derive both the inequality for the excess of work and the inequality for the energy costs from the information exchange fluctuation theorem~\cite{Sagawa2012}.
   This fluctuation theorem has been generalized to for case of multiple systems~\cite{Ito2013}.

 The fluctuation theorem has been also derived for quantum systems~\cite{Morikuni2011,Rana2012,Funo2013}, but our comprehension is still limited.
  In the present article, we present quantum versions of the Jarzynski equalities for the energy costs of the measurement and the information erasure.
 Using the Jensen inequality, we can derive two inequalities which bound the energy costs.
  The inequalities include Sagawa and Ueda's inequalities~\cite{Sagawa2009} as a special case.

   An equality similar to our Jarzynski equality for the energy cost of the measurement was recently derived by Funo, Watanabe and Ueda \cite{Funo2013}. 
   Their formulation, however, assumed that the final state is in equilibrium, whereas our formulation does not.

\section{Memory and Information Process}
  \subsection{System, memory and bath}
  We hereafter focus of isothermal processes. 
  The system which we consider consists of a thermodynamic system $\Sys$, a memory $\Mem$ and a heat bath $\Bath$.
The thermodynamic system $\Sys$ is the target of the measurement.
The memory $\Mem$ stores the information on the outcomes of the measurement.
The heat bath $\Bath$ is at an inverse temperature $\beta$ and is in contact with $\Mem$.

  Let us define the memory $\Mem$ in more details, following Ref.~\cite{Sagawa2009}.
  The memory $\Mem$ is a quantum system in the Hilbert space $\mathcal{H}^\Mem$.
  The space $\mathcal{H}^\Mem$ is divided into $(N+1)$ pieces of mutually orthogonal subspaces $\mathcal{H}^\Mem_a \left( a=0, 1, \dots, N \right)$ as in $\mathcal{H}^\Mem = \oplus_{a=0}^N \mathcal{H}^\Mem_a$.
  The subscript $a$ indicates the result of the measurement; we consider the outcome $a$ to be stored in $\Mem$ when the support of the density operator of $\Mem$ is in the subspace $\mathcal{H}^\Mem_a$. 
  The Hamiltonian of the memory corresponding to $a$ is written as follows:
    \begin{equation}
    \HM_a = \sum_{n_a}^{} \EM{n_a} \PM{n_a}
    \label{},
  \end{equation}
 where $\EM{n_a}$ is an eigenvalue and $\PM{n_a}$ is the projection operator onto the energy eigenstate $\ket{\EM{n_a}}$.
  We refer to the projection operator onto the subspace $\mathcal{H}^\Mem_a$ as $\hPi{\Mem}{a} = \sum_{n_a}^{} \PM{n_a}$.
We refer to the canonical state and the Helmholtz free energy corresponding to the result $a$ at the inverse temperature $\beta$ as
  \begin{align}
    \rhoM{a,\can} &= \frac{e^{-\beta \HM_a}}{Z^\Mem_a}
    = \sum_{n_a}^{} \frac{e^{-\beta \EM{n_a}}}{Z^\Mem_a},\\
    F^\Mem_a &= -\beta^{-1} \log Z^\Mem_a ,
    \label{}
  \end{align}
  respectively, where $Z^\Mem_a = \nTr{e^{ -\beta \HM_a }}$.

  \subsection{Measurement Process}
 We first consider the measurement process from $t=0$ to $t=\tau_\meas$.
 During the process, the thermodynamic system $\Sys$ and the memory $\Mem$ interact with each other under the Hamiltonian $\HSM(t)$ and the memory $\Mem$ and the heat bath $\Bath$ under the Hamiltonian $\HMB(t)$.
 We assume that $\HSM(0) = \HSM(\tau_\meas) = 0$ and $\HMB(0) = \HMB(\tau_\meas) = 0$. 
 We also assume that there is no direct interaction between $\Sys$ and $\Bath$.  
Therefore, the Hamiltonian of the whole system is written as follows:
    \begin{equation}
      \Hami^\tot_\meas(t) = \HS(t) + \HM_\meas(t) + \HB + \HSM(t) + \HMB(t)
      \label{},
    \end{equation}
where $\HS(t)$, $\HM_\meas(t)$ and $\HB$ are the Hamiltonians of $\Sys$, $\Mem$ and $\Bath$, respectively. 
We assume that $\HM_\meas(0) = \HM_\meas(\tau_\meas) = \sum_{a = 0}^{N} \HM_a$.
 
As the initial state, we introduce the density matrix $\rhoSMB{\init}$ of the whole system as follows:
\begin{equation}
\rhoSMB{\init} = \rhoS{\init} \Ox \rhoM{\init} \Ox \rhoB{\can},
\end{equation}
where $\rhoS{\init}$, $\rhoM{\init}$ and $\rhoB{\can}$ are the density matrices of $\Sys$, $\Mem$ and $\Bath$, respectively.
The state $\rhoS{\init}$ is an arbitrary state of the system $\Sys$. 
The state  $\rhoM{\init}$ is the mixture of the canonical distributions of the memory subsystems $\mathcal{H}^\Mem_a$ with non-zero probability $p_\init(a)$;
    \begin{equation}
      \rhoM{\init} = \sum_{a}^{} p_\init(a) \rhoM{a,\can}.
      \label{}
    \end{equation}
    We note that Sagawa and Ueda's inequality~\cite{Sagawa2009} was derived in the case
    \begin{equation}
      p_\init \left( a \right) = \delta_{a,0},
      \label{eq:limit}
    \end{equation}
    where is the error-free limit, whereas we here allow for errors in the initial state of the memory $\Mem$.
The state $\rhoB{\can}$ is the canonical distribution of the bath $\Bath$ with the inverse temperature $\beta$: 
    \begin{equation}
      \rhoB{\can} = \frac{e^{-\beta \HB}}{Z^\Mem} 
      = \sum_{k}^{} \frac{e^{-\beta \EB{k}}}{Z^\Bath} \PB{k}
      \label{eq:Bath_can},
    \end{equation}
     where $Z^\Bath = \nTr{e^{-\beta \HB}}$  and $\EB{k}$ and $\PB{k}$ are the energy eigenvalue and the eigenstate of $\HB$, respectively.
We also define the probability $q^\Sys_\init(i)$ with the orthonormal basis $\set{ \ket{\varphi^\Sys(i)} }$ of $\rhoS{\init}$:
    \begin{equation}
      \rhoS{\init} = \sum_{i}^{} q^\Sys_\init(i) \ketbra{\varphi^\Sys(i)}.
      \label{}
    \end{equation}

\begin{figure}
		\centering
		\includegraphics[height=17cm, clip]{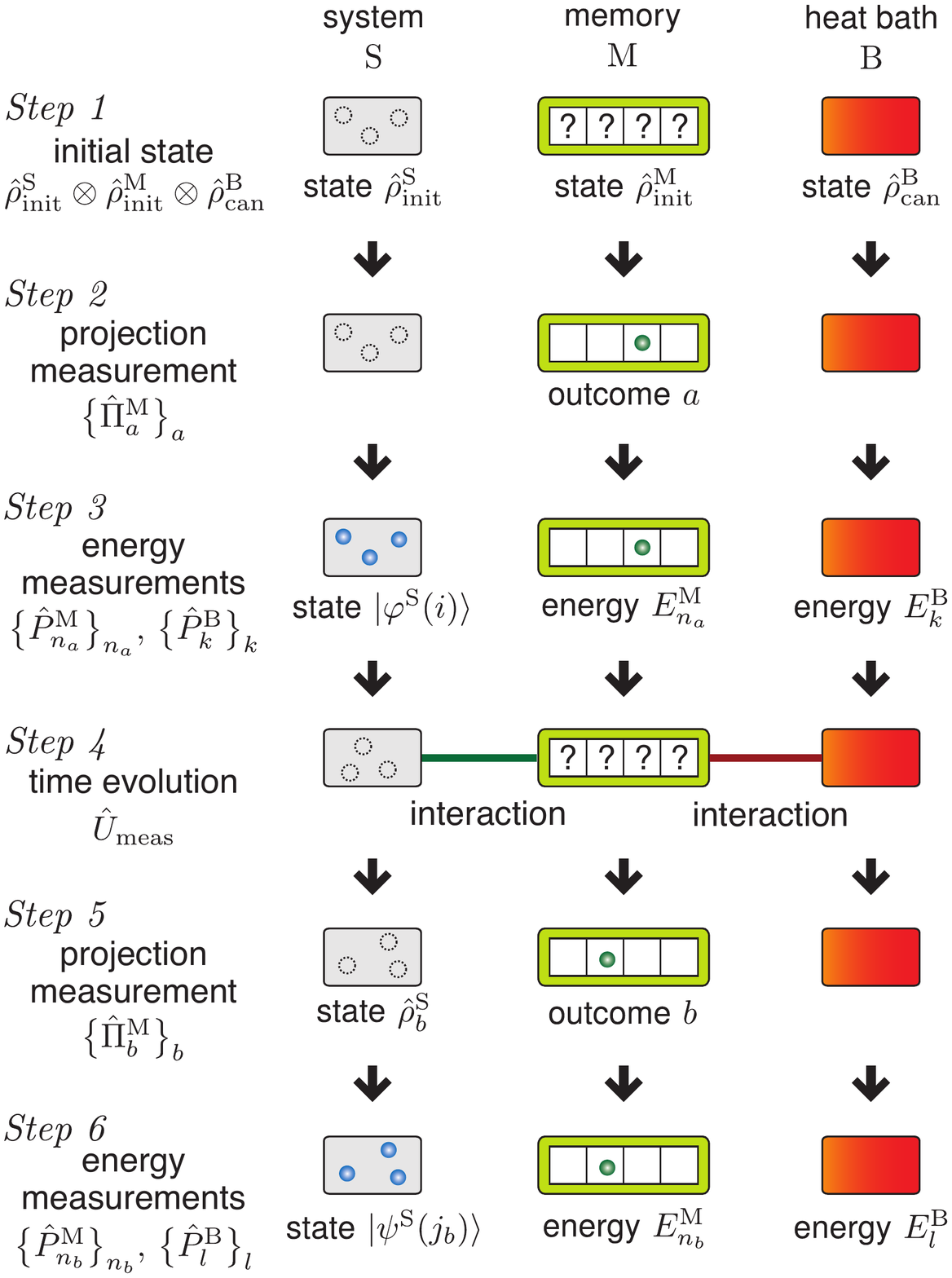}
		\caption{(Color online) Schematic illustration of the measurement process.
				(\textit{Step 1}) The initial state of the whole state is given by $\rhoS{\init} \Ox \rhoM{\init} \Ox \rhoB{\can}$. 
				(\textit{Step 2}) We perform a projection measurement $\set{\hPi{\Mem}{a}}_a$ on $\Mem$ and obtain the outcome $a$. 
				(\textit{Step 3}) We measure the energy of $\Mem$ and $\Bath$ using the projection measurement $\set{\PM{n_a}}_{n_a}$ and $\set{\PB{k}}_k$ and obtain the energy $\EM{n_a}$ and $\EB{k}$, respectively.
				We refer to the state of $\Sys$ as $\ket{\varphi^\Sys(i)}$.
				(\textit{Step 4}) We perform the time evolution $\hU_\meas$.
				(\textit{Step 5}) We perform a projection measurement $\set{\hPi{\Mem}{b}}_b$ on $\Mem$ and obtain the measurement outcome $b$.
				Then the density matrix of $\Sys$ is given by $\rhoS{b}$.
				(\textit{Step 6}) We measure the energy of $\Mem$ and $\Bath$ using the projection measurement $\set{\PM{n_b}}_{n_b}$ and $\set{\PB{l}}_l$, respectively.
				We obtain the energy $\EM{n_b}$ and $\EB{l}$, respectively.
				We refer to the state of $\Sys$ as $\ket{\psi^\Sys(j_b)}$.
		}
		\label{fig:meas_process}
\end{figure}

As the measurement process, we perform the following steps (see Fig.~\ref{fig:meas_process}): 
    \Step{1}
      We prepare the initial state $\rhoSMB{\init}$ of the whole state.
    \Step{2}
      We fix the state of the memory $\Mem$ by performing a projection measurement $\set{ \hPi{\Mem}{a} }_a$ on $\Mem$.
      We refer to the result of the measurement as $a$.
    \Step{3}
      We measure the energy of $\Mem$ and $\Bath$ by performing the projection measurements $\set{\PM{n_a}}_{n_a}$ and $\set{\PB{k}}_k$ and refer to the results of the measurements as $n_a$ and $k$, respectively.
      We also refer to the state of the system $\Sys$ at this timing as $\ket{\varphi^\Sys(i)}$.
    \Step{4}
      We perform the time evolution 
      \begin{equation}
        \hU_\meas = \mathcal{T} \exp \left( -\frac{i}{\hbar} \int_{0}^{\tau_\meas} \Hami^\tot_\meas(t) dt \right),
        \label{}
      \end{equation}
      where $\mathcal{T}$ is the time-ordering operator.
    \Step{5}
      We perform a projection measurement $\set{\hPi{\Mem}{b}}_b$ on $\Mem$ and refer to the result of the measurement as $b$.
      Then, the density matrix of $\Sys$ is given by
      \begin{equation}
        \rhoS{b} = \frac{\sTr{\Mem\Bath}{\hPi{\Mem}{b} \hU_\meas \rhoSMB{\init} \hU_\meas^\dg \hPi{\Mem}{b}}}{\Tr{\hPi{\Mem}{b} \hU_\meas \rhoSMB{\init} \hU_\meas^\dg \hPi{\Mem}{b}}}.
        \label{}
      \end{equation}
      We refer to the eigenvalue and eigenvector of the density matrix $\rhoS{b}$ as $q^\Sys_\meas(j_b)$ and $\ket{\psi^\Sys(j_b)}$, respectively.
    \Step{6}
      We measure the energy of $\Mem$ and $\Bath$ by performing the projection measurements $\set{\PM{n_b}}_{n_b}$ and $\set{\PB{l}}_l$ and refer to the results of the measurement as $n_b$ and $l$, respectively.
      We also refer to the state of $\Sys$ at this timing as $\ket{\psi^\Sys(j_b)}$, which is an eigenvector of $\rhoS{b}$.
      
      In the process above, the probability that the initial and final states of the system $\Sys$ are $\ket{\varphi^\Sys(i)}$ and $\ket{\psi^\Sys(j_b)}$, respectively, and that the results of the all measurements are $(a, n_a, k, b, n_b, l)$ is written as follows;
      \begin{multline}
        p_\all \left( i, j_b, a, n_a, k, b, n_b, l \right)           
          = q^\Sys_\init(i) p_\init(a) \frac{e^{-\beta \EM{n_a}}}{Z^\Mem_a}
          \frac{e^{-\beta \EB{k}}}{Z^\Bath} \\
          \times \,
          \lTr{\left( \QS{j_b} \Ox \PM{n_b} \Ox \PB{l} \right)
              \hU_\meas
              \left( \QS{i,\init} \Ox \PM{n_a} \Ox \PB{k} \right)
              \hU_\meas^\dg
              \left( \QS{j_b} \Ox \PM{n_b} \Ox \PB{l} \right)
              } ,
        \label{}
      \end{multline}
     where $\QS{i,\init} \equiv \ketbra{\varphi^\Sys(i)}$ and $\QS{j_b} \equiv \ketbra{\psi^\Sys(j_b)}$.
      Thus, the probability of the result of measurement  $\set{\hPi{\Mem}{b}}_b$ is 
      \begin{equation}
        p_\meas(b)
          = \sum_{i, j_b, a, n_a, k, n_b, l}^{}
                    p_\all \left(i, j_b, a, n_a, k, b, n_b, l \right)
          =	\Tr{\hPi{\Mem}{b} \hU_\meas \rhoSMB{\init} \hU_\meas^\dg \hPi{\Mem}{b}}.
        \label{}
      \end{equation}
      We also refer to the expectation value of an arbitrary function $f(i, j_b, a, n_a, k, b, n_b, l)$ as
      \begin{equation}
        \mavr{f} = \sum_{i, j_b, a, n_a, k, b, n_b, l}^{}
            p_\all \left(i, j_b, a, n_a, k, b, n_b, l \right)
            f(i, j_b, a, n_a, k, b, n_b, l).
        \label{}
      \end{equation}

    In the process, the work $\Wm$ done on $\Mem$ and the difference $\Delta\Fm$ of the Helmholtz free energy are
    \begin{align}
      \Wm(n_a,k,n_b,l) &= \left( \EM{n_b} + \EB{l} \right) - \left( \EM{n_a} + \EB{k} \right), \\
      \Delta\Fm(a,b) &= F^\Mem_b - F^\Mem_a,
    \end{align}
    respectively.
    We define the information gain $\Delta\Hm$  of $\Mem$ and the information loss $I$ of $\Sys$ as
    \begin{align}
      \Delta\Hm(a,b) &= -\log p_\meas(b) + \log p_\init(a) , \\
      I(i,j_b) &= -\log q^\Sys_\init(i) + \log q^\Sys_\meas(j_b) ,
    \end{align}
    respectively.
    We can write the expectation value of $\Delta\Hm$ as the difference of the Shannon entropy $H(\bm{p}) = -\sum_x p(x)\log p(x)$:
    \begin{equation}
    	\mavr{\Delta\Hm} = H(\bm{p}_\meas) - H(\bm{p}_\init).
    \end{equation}
    We can also write the expectation value of $I$ as
    \begin{equation}
      \mavr{I} = S\left( \rhoS{\init} \right) - \sum_{b}^{} p_\meas(b) S\left( \rhoS{b} \right)
      \label{}, 
    \end{equation}
    where $S(\hrho{}{})$ is the von Neumann entropy $S(\hrho{}{}) = -\Tr{\hrho{}{}\log\hrho{}{}}$.
    This is equal to the QC-mutual information~\cite{Groenewold1971,Ozawa1986,Sagawa2008}.

  \subsection{Erasure Process}
  Next, we consider the erasure process from $t=0$ to $t=\tau_\eras$.
  	In the present process, we perform information erasure of the memory $\Mem$ using the interaction between $\Mem$ and $\Bath$.

    The interaction between $\Mem$ and $\Bath$ follows the Hamiltonian $\hMB(t)$.
     We assume that $\hMB(0) = \hMB(\tau_\eras) = 0$.
    The Hamiltonian of the whole system is written as 
    \begin{equation}
      \Hami^\tot_\eras(t) = \HM_\eras(t) + \HB + \hMB(t)
      \label{},
    \end{equation}
    where $\HM_\eras(t)$ and $\HB$ are the Hamiltonians of $\Mem$ and $\Bath$, respectively.
    We assume that $\HM_\eras(0) = \HM_\eras(\tau_\eras) = \sum_{b = 0}^{N} \HM_b$.

    As the initial state of the whole system, we introduce 
   \begin{equation}
   \rhoMB{\meas} =\rhoM{\meas} \Ox \rhoB{\can},
   \end{equation}
   where $\rhoB{\can}$ is the one in \eqref{eq:Bath_can} and $\rhoM{\init}$ is a mixture of the canonical distributions $\rhoM{b,\can}$ with non-zero probability $\Tilde{p}_\meas(b)$;
    \begin{equation}
      \rhoM{\meas} = \sum_{b}^{} \Tilde{p}_\meas(b) \rhoM{b,\can}.
      \label{}
    \end{equation}
    We here used the subscript $\meas$ for the initial states because we suppose that the information erasure process takes place after the measurement process.
    
    \begin{figure}
		\centering
		\includegraphics[height=17cm, clip]{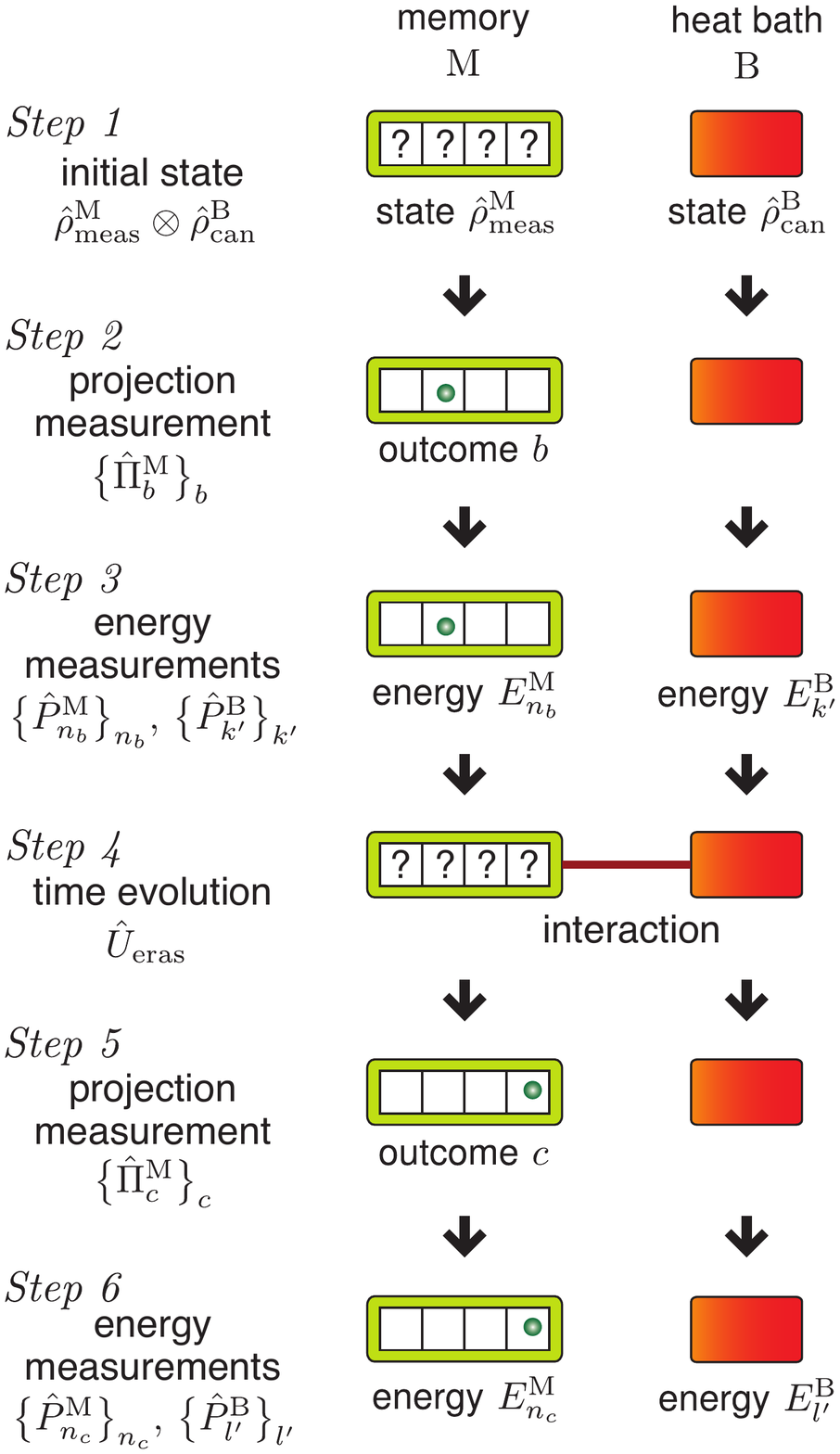}
		\caption{(Color online) Schematic illustration of the erasure process.
				(\textit{Step 1}) The initial state of the whole state is given by $\rhoM{\meas} \Ox \rhoB{\can}$. 
				(\textit{Step 2}) We perform a projection measurement $\set{\hPi{\Mem}{b}}_b$ on $\Mem$ and obtain the outcome $b$.
				(\textit{Step 3}) We measure the energy of $\Mem$ and $\Bath$ using the projection measurement $\set{\PM{n_b}}_{n_b}$ and $\set{\PB{k^\prime}}_{k^\prime}$ and obtain the energy $\EM{n_b}$ and $\EB{k^\prime}$, respectively.
				(\textit{Step 4}) We perform the time evolution $\hU_\eras$.
				(\textit{Step 5}) We perform a projection measurement $\set{\hPi{\Mem}{c}}_c$ on $\Mem$ and obtain the outcome $c$.
				(\textit{Step 6}) We measure the energy of $\Mem$ and $\Bath$ using the projection measurement $\set{\PM{n_c}}_{n_c}$ and $\set{\PB{l^\prime}}_{l^\prime}$ and obtain the energy $\EM{n_c}$ and $\EB{l^\prime}$, respectively.
		}
		\label{fig:eras_process}
\end{figure}
    
    We perform the following information erasure process (see Fig.~\ref{fig:eras_process}):
    \Step{1}
      We introduce $\rhoMB{\meas}$ as the initial state of the whole system.
    \Step{2}
      We fix the state of the memory $\Mem$ initially by performing a projection measurement $\set{ \hPi{\Mem}{b} }_b$ on $\Mem$.
      We refer to the result of the measurement as $b$.
    \Step{3}
      We measure the energy of $\Mem$ and $\Bath$ by performing the projection measurements $\set{\PM{n_b}}_{n_b}$ and $\set{\PB{k^\prime}}_{k^\prime}$ and refer to the results of the measurements  $n_a$ and $k^\prime$, respectively.
    \Step{4}
      We perform the time evolution which is written as 
      \begin{equation}
        \hU_\eras = \mathcal{T} \exp \left( -\frac{i}{\hbar} \int_{0}^{\tau_\eras} \Hami^\tot_\eras(t) dt \right).
        \label{}
      \end{equation}
    \Step{5}
      We perform a projection measurement $\set{\hPi{\Mem}{c}}_c$ on $\Mem$ and refer to the result of the measurement as $c$.
    \Step{6}
      We measure the energy of $\Mem$ and $\Bath$ by performing the projection measurements $\set{\PM{n_c}}_{n_c}$ and $\set{\PB{l^\prime}}_{l^\prime}$ and refer to the results of the measurements as $n_c$ and $l^\prime$, respectively.

      The probability that the whole set of the results of the measurements is $(b, n_b, k^\prime, c, n_c, l^\prime)$ is 
        \begin{equation}
          \Tilde{p}_\all \left(a, n_a, k^\prime, b, n_b, l^\prime \right) 
          = \Tilde{p}_\meas(b) \frac{e^{-\beta \EM{n_b}}}{Z^\Mem_b}
            \frac{e^{-\beta \EB{k^\prime}}}{Z^\Bath}
            \lTr{\left(\PM{n_c} \Ox \PB{l^\prime} \right)
            		\hU_\eras
								\left( \PM{n_b} \Ox \PB{k^\prime} \right)
								\hU_\eras^\dg
								\left( \PM{n_c} \Ox \PB{l^\prime} \right)
								}.
          \label{}
        \end{equation}
  	    Thus, the probability that we obtain the result $c$ is
          \begin{equation}
            p_\eras(c)
						= \sum_{b, n_b, k^\prime, n_c, l^\prime}^{}
										p_\all \left(b, n_b, k^\prime, c, n_c, l^\prime \right)
						=	\Tr{\hPi{\Mem}{c} \hU_\eras \rhoMB{\meas} \hU_\eras^\dg \hPi{\Mem}{c}}
					\label{}.
      	\end{equation}
				We refer to the expectation value of an arbitrary function $g(b, n_b, k^\prime, c, n_c, l^\prime)$ as
				\begin{equation}
					\eavr{g} = \sum_{b, n_b, k^\prime, c, n_c, l^\prime}^{}
							\Tilde{p}_\all \left(b, n_b, k^\prime, c, n_c, l^\prime \right)
							g(b, n_b, k^\prime, c, n_c, l^\prime)
					\label{}.
				\end{equation}
	
      We define the work $\We$ done on the memory $\Mem$, the difference of the free energy $\Delta\Fe$ and the information gain $\Delta\He$ of $\Mem$ during the information erasure process as
    \begin{align}
      \We(n_b,k^\prime,n_c,l^\prime) &= \left( \EM{n_c} + \EB{l^\prime} \right) - \left( \EM{n_b} + \EB{k^\prime} \right) , \\
      \Delta\Fe(b,c) &= F^\Mem_c - F^\Mem_b , \\
      \Delta\He(b,c) &= -\log p_\eras(c) + \log \Tilde{p}_\meas(b).
    \end{align}
         
\section{Jarzynski Equalities}
	For the above measurement and information erasure processes, we derive two quantum versions of the Jarzynski equality:
	\begin{gather} 
		\mavr{e^{-\beta (\Wm - \Delta \Fm) - \Delta \Hm + I}} = 1,
	  \label{eq:meas_je}\\
		\eavr{e^{-\beta (\We - \Delta \Fe) - \Delta \He}} = 1.
		\label{eq:eras_je}
	\end{gather}
	
  \textbf{Proof}
    We first derive \eqref{eq:meas_je}.  
    Using $\sum_i \QS{i,\init} = \idop^\Sys$, $\sum_a \sum_{n_a} \PM{n_a} = \idop^\Mem$ and  $\sum_k \PB{k} = \idop^\Bath $, where $\idop^\Sys$, $\idop^\Mem$ and $\idop^\Bath$ are the identity operators for $\Sys$, $\Mem$ and $\Bath$, respectively, we obtain \eqref{eq:meas_je} as follows:
    \begin{align}
      &\mavr{e^{-\beta (\Wm - \Delta \Fm) - \Delta \Hm + I}} \notag\\
      &= \sum_{i, j_b, a, n_a, k, b, n_b, l}^{} 
            q^\Sys_\meas(j_b) p_\meas(b) \frac{e^{-\beta \EM{n_b}}}{Z^\Mem_b}
            \frac{e^{-\beta \EB{l}}}{Z^\Bath} \notag\\
            & \qquad\qquad \times \,
            \lTr{\left( \QS{j_b} \Ox \PM{n_b} \Ox \PB{l} \right)
                \hU_\meas
                \left( \QS{i,\init} \Ox \PM{n_a} \Ox \PB{k} \right)
                \hU_\meas^\dg
                \left( \QS{j_b} \Ox \PM{n_b} \Ox \PB{l} \right)
                } \notag \\
      &= \sum_{ j_b, k, b, n_b, l}^{}
         q^\Sys_\meas(j_b) p_\meas(b) \frac{e^{-\beta \EM{n_b}}}{Z^\Mem_b}
            \frac{e^{-\beta \EB{l}}}{Z^\Bath}
            \Tr{\QS{j_b} \Ox \PM{n_b} \Ox \PB{l} } \notag \\
      &= \sum_b\, p_\meas(b) \Tr{ \rhoS{b} \Ox \rhoM{b,\can} \Ox \rhoB{\can}} \notag \\
      &= 1.
    \end{align}
    We can also derive \eqref{eq:eras_je} in a way similar to the above:
    \begin{align}
      &\eavr{e^{-\beta (\We - \Delta \Fe) - \Delta \He}} \notag\\  
      &= \sum_{b, n_b, k^\prime, c, n_c, l^\prime}^{}
         p_\eras(c) \frac{e^{-\beta \EM{n_c}}}{Z^\Mem_c}
              \frac{e^{-\beta \EB{l^\prime}}}{Z^\Bath}
              \Tr{\left(\PM{n_c} \Ox \PB{l^\prime} \right)
                  \hU_\eras
                  \left( \PM{n_b} \Ox \PB{k^\prime} \right)
                  \hU_\eras^\dg
                  \left( \PM{n_c} \Ox \PB{l^\prime} \right)
                  }  \notag \\
      &= \sum_{c, n_c, l^\prime}^{}
         p_\eras(c) \frac{e^{-\beta \EM{n_c}}}{Z^\Mem_c}
              \frac{e^{-\beta \EB{l^\prime}}}{Z^\Bath}
              \Tr{\PM{n_c} \Ox \PB{l^\prime} }  \notag \\
      &= \sum_{c}^{}\,
         p_\eras(c) \Tr{\rhoM{c,\can} \Ox \rhoB{\can} }  \notag \\
      &= 1.
    \end{align}	
    {\hfill$\Box$}
	
	By using the Jensen inequality $\avr{e^f} \geq e^{\avr{f}}$, we can reduce Eqs.~\eqref{eq:meas_je} and \eqref{eq:eras_je} into the inequalities for the energy costs of the measurement process and the information erase process;
	\begin{gather}
		\mavr{\Wm} \geq \mavr{\Delta\Fm} -\beta^{-1} \left( \mavr{\Delta\Hm} - \mavr{I} \right) ,
		\label{eq:meas_ineq}\\
		\eavr{\We} \geq \eavr{\Delta\Fe} -\beta^{-1} \eavr{\Delta\He}
		\label{eq:eras_ineq}.
	\end{gather}
	
  Let us consider the error-free limit~\eqref{eq:limit}; in other words, we fix the state of the memory $\Mem$ to $a$ in the initial state.
  Then, the average $\mavr{\Delta\Hm}$ is reduced to $H(\bm{p}_\meas)$ and the inequality \eqref{eq:meas_ineq} is reduced to
  \begin{equation}
  	\mavr{\Wm} \geq \mavr{\Delta\Fm} -\beta^{-1} \left( H(\bm{p}_\meas) - \mavr{I} \right),
  \end{equation}
  which is equivalent to Sagawa and Ueda's inequality for the energy cost of measurement~\cite{Sagawa2009}.
  Similarly, the inequality \eqref{eq:eras_ineq} is reduced to Sagawa and Ueda's inequality for the energy cost of the information erasure~\cite{Sagawa2009} in the limit of $p_\eras(c) \to \delta_{c,0}$; 
  \begin{align}
  	\eavr{\Delta\Fe} &\to -\mavr{\Delta\Fm}, \\
	  \eavr{\Delta\He} &\to -H(\Tilde{\bm{p}}_\meas).
	\end{align}
  Thus, our results are generalization of Sagawa and Ueda's inequalities for the energy costs.
  

\section{Conclusion}
	We obtain two quantum versions of the Jarzynski equality for the information processes;
	the first one is for the energy cost of the measurement process, whereas the second one is for the energy cost of the information erasure.
	Using the Jensen inequality, we can reduce these Jarzynski equalities into inequalities for the energy costs of the measurement process and the information erasure process.
	The inequalities include Sagawa and Ueda's inequalities for the energy costs as a special case of the error-free limit.
	With this result and the result in Ref.~\cite{Morikuni2011}, we can treat both of the two types of researches as in the form of fluctuation theorem.
	In the present formulation, we cannot treat the case in which the system and the memory initially have quantum correlation.
  It is a good future problem, which would give a quantum counterpart of the one in Ref.~\cite{Sagawa2012}.

\bibliographystyle{apsrev4-1}
%

\end{document}